\documentclass{article}
\usepackage{spconf,amsmath,graphicx,tipa,subfigure,color}
\usepackage{amssymb,booktabs}
\usepackage{url}
\usepackage{cite}
\usepackage{multirow}
\usepackage[table,xcdraw]{xcolor}
\usepackage[T1]{fontenc}
\usepackage{newtxmath,newtxtext}

\title{R-G2P: EVALUATING AND ENHANCING ROBUSTNESS OF GRAPHEME TO PHONEME CONVERSION BY CONTROLLED NOISE INTRODUCING AND CONTEXTUAL INFORMATION INCORPORATION
}

\name{Chendong Zhao$^{\star \dagger \ddagger}$\thanks{$\ddagger$ Work done during an internship at Ping An Technology. } \qquad Jianzong Wang$^{\dagger * }$\thanks{* Corresponding author: Jianzong Wang, jzwang@188.com.} \qquad Xiaoyang Qu$^{\dagger}$ \qquad Haoqian Wang$^{\star}$ \qquad Jing Xiao$^{\dagger}$ }

\address{ $^{\dagger}$ Ping An Technology (Shenzhen) Co., Ltd. \\
$^{\star}$ The Shenzhen International Graduate School, Tsinghua University, China}

\begin{document}

\maketitle

\begin{abstract}
Grapheme-to-phoneme (G2P) conversion is the process of converting the written form of words to their pronunciations. It has an important role for text-to-speech (TTS) synthesis and automatic speech recognition (ASR) systems. In this paper, we aim to evaluate and enhance the robustness of G2P models.
We show that neural G2P models are extremely sensitive to orthographical variations in graphemes like spelling mistakes.
To solve this problem, we propose three controlled noise introducing methods to synthesize noisy training data.
Moreover, we incorporate the contextual information with the baseline and propose a robust training strategy to stabilize the training process. 
The experimental results demonstrate that our proposed robust G2P model (r-G2P) outperforms the baseline
significantly (-2.73\% WER on Dict-based benchmarks and -9.09\% WER on Real-world sources).
\end{abstract}


\section{Introduction}
\label{sec:intro}
Grapheme-to-phoneme (G2P) conversion generates a phonetic transcription from the written form of words. It is essential to develop a phonemic lexicon for TTS and ASR systems~\cite{1,2,3,4}. For this purpose, G2P techniques are used.
For instance, modern TTS systems adopt G2P models as their frontend. Thus the performance of the overall system depends on the accuracy of G2P conversion.
Meanwhile, TTS systems need to pronounce real-world inputs of diverse sources, like web pages or translated texts.
The noise of real-world data can exhibit in many forms, such as spelling mistakes, newly emerged words, or even changed spellings over time.
Unlike human who can easily pronounce these, G2P models may fail badly on such words.
For example, "occurred" is commonly misspelled as "occured".
The Transformer G2P~\cite{trans} converts it to [\textschwa kj\textupsilon r\textlengthmark d], which should be [\textschwa k\textepsilon \textlengthmark d]. "pronounciation" is another common misspelling with an extra 'o', this typo results in [pro\textupsilon n\textscripta \textupsilon nsi\textlengthmark e\i \textesh \textschwa n], which should be [pro\textupsilon n\textschwa nsi\textlengthmark e\i \textesh \textschwa n].

G2P models have shown well performances on the clean text~\cite{ic1,ic2}, but they suffer from substandard real-world inputs.
Indeed, G2P conversion can be considered as a neural machine translation (NMT) task, where we need to translate source graphemes into target phonemes.
While many researches have focused on the robustness of NMT~\cite{adn}, this is the first study that evaluates and enhances the robustness of G2P conversion in real-world noisy scenarios.
One of state-of-the-art models, Transformer G2P, is adopted as our baseline.
Our contributions are three-fold:
\vspace{-2mm}
\begin{itemize}
\item[$\bullet$] We confirm G2P models' vulnerability to real-world noise, then statistically analyze the conversion failures. We define three noise types based on the statistics.
\vspace{-3mm}
\item[$\bullet$] We investigate three controlled noise introducing methods to synthesis our training data, which are proven to be effective in the experiments.
\vspace{-3mm}
\item[$\bullet$] We integrate and investigate the contextural information with the Transformer G2P, and propose a robust training strategy to mitigate the effect of noise during training.
\end{itemize}

\section{Real-World Noise Effect }
\label{sec:real}
\begin{figure}[t]
	\centering
	\includegraphics[width=1.0\linewidth]{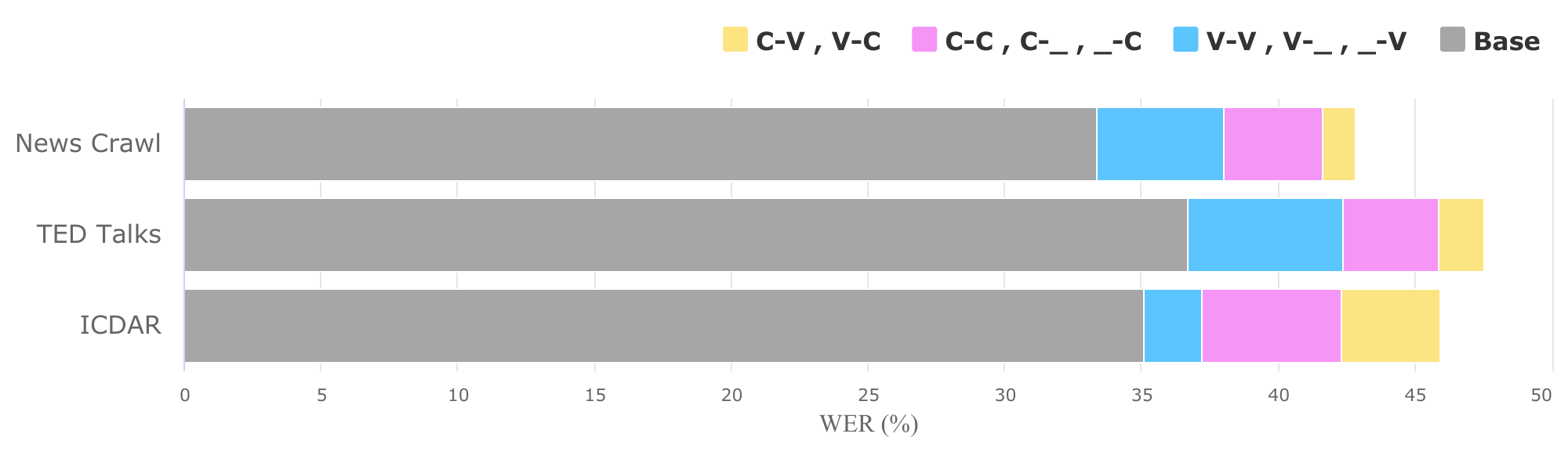}
	\vspace{-6mm}
	\caption{ Conversion failures on multimodal sources. \textbf{Base} denotes that the text is correct but wrongly converted by the baseline. Other failures are caused by the noise, including vowel noise (in blue), constant noise (in pink) and substitutions of vowels and constants (in yellow). Take the vowel noise for example, V-V means that the failure is caused by a vowel being substituted by another vowel ("than" -> "then"). V-\_ means there's a vowel deletion noise ("neighbour" -> "neighbor") while \_-V means a vowel insertion noise ("lose" -> "loose").}
	\label{fig:err}
\end{figure}

In this Section, we study the effect of real-world noise on the baseline performance.
The goals of this section are: (i) evaluating the robustness of the baseline across multimodal sources, (ii) statistically analyzing the noise distributions, and (iii) providing statistics to craft a noisy training dataset.
We analyze the noise in three aspects.
First, there are three major sources for G2P, including direct texts (e.g. web-crawled news), raw transcriptions from ASR, and recognitions from optical character recognition (OCR) systems.
We adopt News Crawl~\cite{news}, TED Talks~\cite{ted} and ICDAR Post-OCR text correction~\cite{OCR} official dev sets as representative data which contain machine-generated noisy texts.
Second, to detail into phonetic analyses, we group graphemes into \textbf{Vowels} and \textbf{Constants}.
Third, there are three orthographical noise types in each group: \textbf{Insertion}, \textbf{Deletion}, and \textbf{Substitution}.
Overall, we classify the reasons for G2P conversion failures (measured in word error rate) into 4 types, as shown in Fig.~\ref{fig:err}.
The average increased WER caused by 3 noise types are +4.6\% for vowel noise, +4.9\% for constant noise, and +2.6\% for substitutions of constants and vowels. Conversion failures caused by the noise account for almost 1 / 4 of all failures. This shows a great improvement space for G2P models to tackle real-world noise.

\section{Approach}
\label{sec:app}

When converting a noisy word, G2P models would fail easily because the noisy grapheme pattern is not observed frequently enough in the training data.
Further, other neighboring graphemes would be affected by this "less jointly trained" unexpected noise. We refer this phenomenon as "noise propagation". 
To address these problems, we first explore introducing carefully crafted noise in the training data, then regularize the effects of noise by the contextual information incorporation.

\subsection{Controlled noise introducing}

Ideally, we would train a G2P model on parallel data with noisy inputs. Since no such data exists, we propose three controlled noise introducing methods.
For \textbf{nat} and \textbf{syn}, each word is modified with the probability of $p$.

\noindent \textbf{Orthographical natural noise introduction: nat} ~We first explore introducing with weak noise, i.e., the natural noise.
Natural noise denotes that the noisy word has similar pronunciations with the correct one but is orthographically different.
We adopt the Wikipedia dataset~\cite{pedia}
to replace correct words with their misspelled versions.
It is considered weak because it may occur when one is not sure about the spelling form but aware of the pronunciation, so this type of noise wouldn't lead to big changes in the G2P conversion.

\begin{figure}[t]
	\centering
	\includegraphics[width=0.7\linewidth]{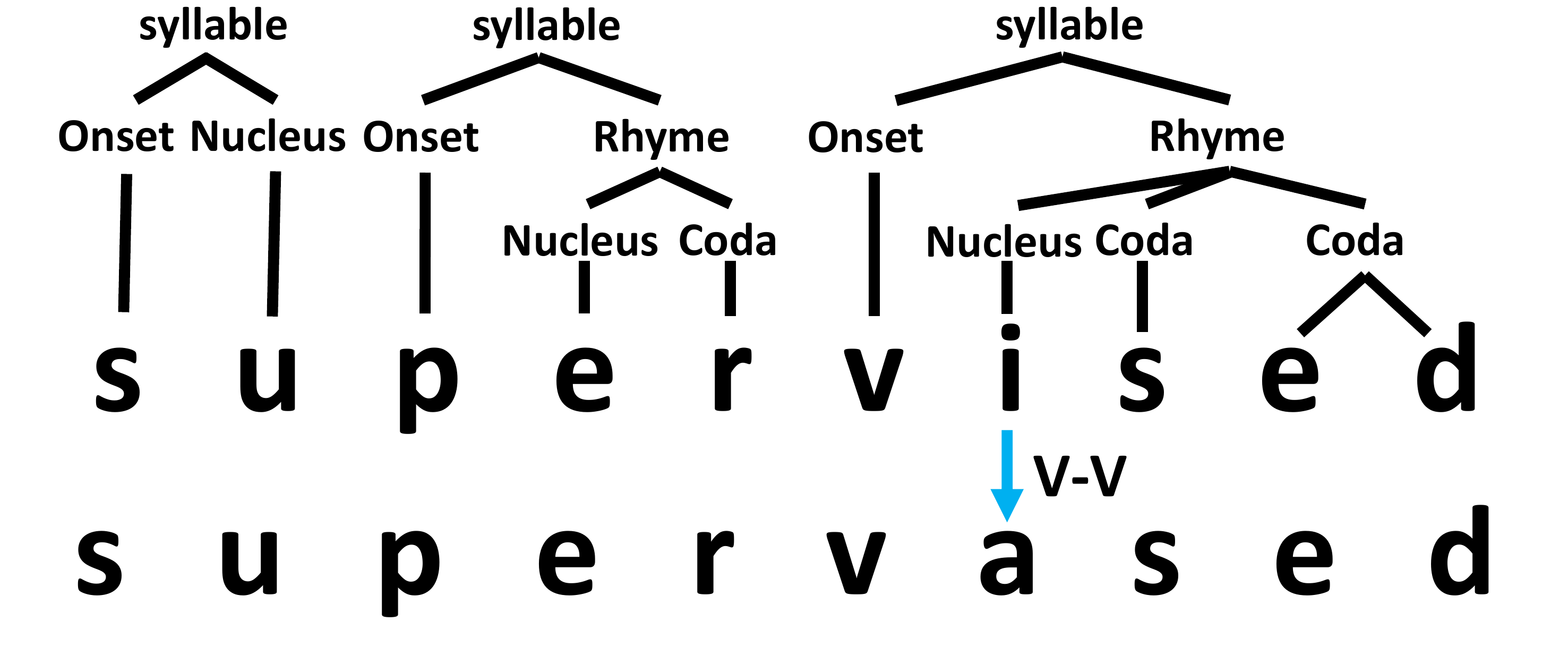}
	\vspace{-2mm}
	\caption{ The illustration of a synthetic noise example.}
	\label{fig:parse}
\end{figure}

\noindent \textbf{Phonological                                                                                                                                                                                                                                                                                                                                                                                                                                                                                                                                                                                                                                                                                                                                                                                                                                                                                                                                                                                                                                                                                                                                                                                                                                                                                                                                                                                                                                                                                                                                                                                                                                                                                                                                                                                                                                                                                                                                                                                                                                                                                                                                                                                                                                                                                                                                                                                                                                                                                                                                                                                                                                                                                                                                                                                                                                                                                                                                                                                                                                                                                                                                                                                                                                                                                                                                                                                                                                                                                                                                                                                                                                                                                                                                                                                                                                                                                                                                                                                                                                                                                                                                                                                                                                                                                                                                                                                                                                                                                                                                                                                                                                                                                                                                                                                                                                                                                                                                                                                                                                                                                                                                                                                                                                                                                                                                                                                                                                                                                                                                                                                                                                                                                                                                                                                                                                                                                                                                                                                                                                                                                                                                                                                                                                                                                                                                                                                                                                                                                                                                                                                                                                                                    noise synthesis: syn} ~Uncareful noise syntheses would easily lead to high-complexity and non-convergence for neural models. 
So,
we synthesis noisy examples basing on the phonology knowledge and noise distributions.
First, as shown in Fig.~\ref{fig:parse}, graphemes could be phonologically categorized into sound units.
These units can further parse into phonetic structures, i.e., syllables.
In practice, we obtain syllables using Consonant Cluster-Vowel approach~\cite{cvv}.
Second, we limit the modifying position within the syllable boundary. This would help to constrain the "noise propagation" within the syllable boundary as well.
Third, we sample and apply one of the 3 noise types in Sec.~\ref{sec:real} to introduce with. The sampling probability
is based on the collected noise distributions in Sec.~\ref{sec:real}.
This 3-step noise synthesis process aims to mimic the real-world data, thus provides a realistic noisy dataset. 

\noindent \textbf{Gradient-based adversarial perturbation: adv}
~The adversarial perturbation strategy has been applied to text classification~\cite{adv} to improve robustness. Motivated by this strategy, we investigate adding the adversarial perturbation to the grapheme embedding.
Let $x , y \in \mathbb{R}^{d \times 1} $ denote the grapheme embedding and phoneme embedding, respectively.
$\hat{\theta}$ and $\theta$ are the last and current iteration model parameters.
During every training iteration, the worst case perturbation $\delta_{ap}$ is added:
\begin{equation}
	\delta_{ap}=\underset{\vert\vert\delta\vert\vert \leq \epsilon}{\arg \min }\log P\left(y \mid x+\delta ; \theta\right),
\end{equation}
where $\epsilon$ is a hyper-parameter to control the magnitude of perturbation. Actually, it is intractable to calculate the minimum of objective function as shown in Eq. 1. In paractice, $\delta_{ap}$ is approximated via the gradient of objective function:
\begin{equation}
	\delta_{ap}=-\epsilon g_{x} / \vert\vert g_{x}\vert\vert_{2}, \text { where } g_{x}=\nabla_{x}\log P\left(y \mid x ; \hat{\theta}\right)
\end{equation}
where $g_x$ denotes the gradient.
The adversarial perturbation is actually a strong regularizer in the high-dimensional space, which substantially increases the diversity of inputs.

\subsection{Incorporating contextual information }
\begin{figure}[t]
	\centering
	\vspace{-1mm}
	\includegraphics[width=1.0\linewidth]{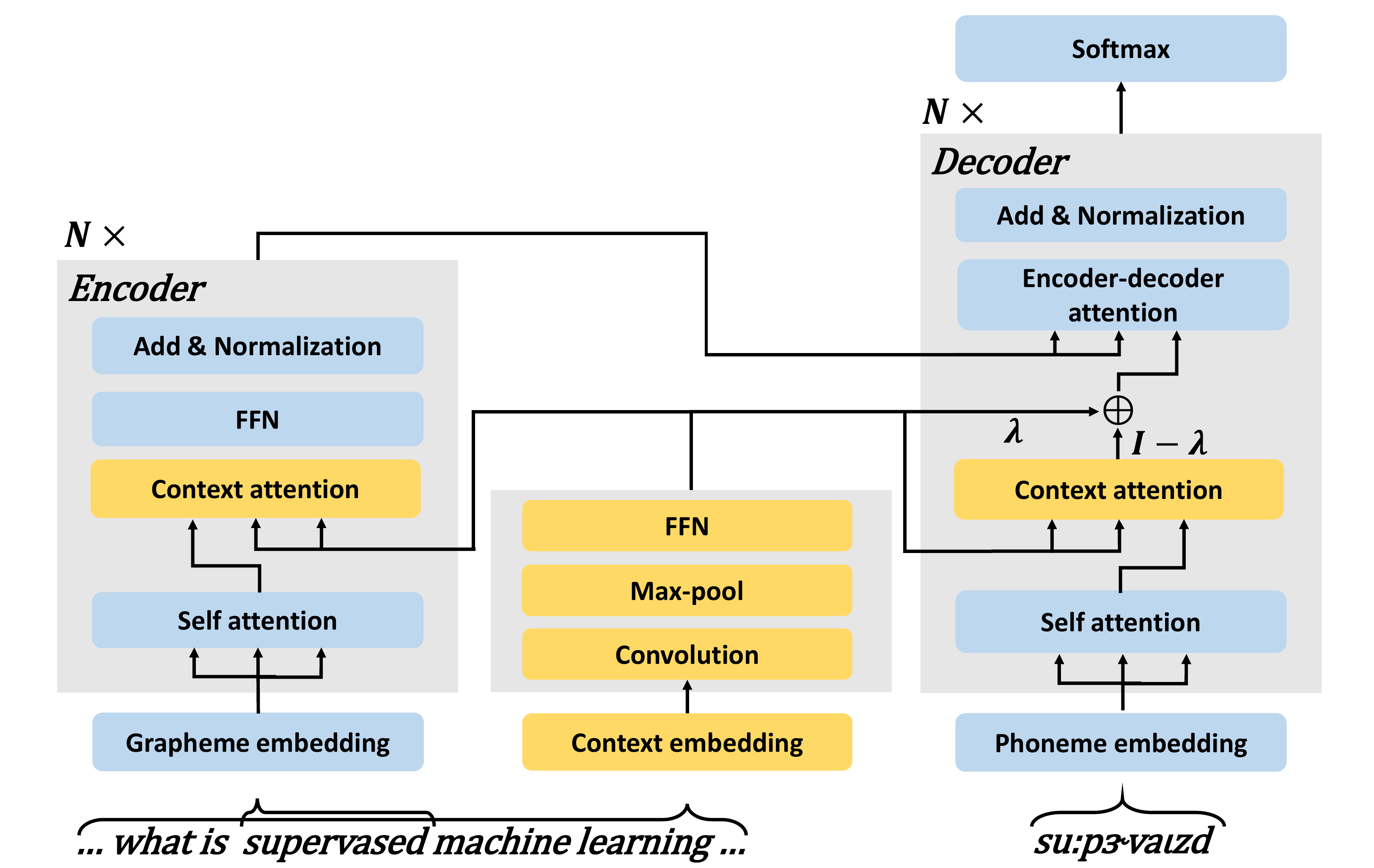}
	\caption{ Our extended Transformer
		G2P model which exploits the contextual information. The original structure is in blue, newly introduced modules are highlighted in yellow.}
	\label{fig:model}
\end{figure}

We incorporate the contextual information to denoise and mitigate the "noise propagation" problem.
To create context ${C}_{k}$ for $k$-th word ${x_k}$, we simply choose $l$ words both left and right.
As shown in Fig.~\ref{fig:model}, we use a convolution layer to compute the context representation.
The convolution layer aims to traverse the context embedding in turn, thus extracts feature vectors.

We integrate ${C}_{k}$ into both the encoder and decoder of the original Transformer.
In the encoder, let $A_k$ be the output of the self-attention. The second sublayer integrates the context:
\begin{equation}
	{M}_{k}=\operatorname{MultiHead}\left({A}_{k}, {C}_{k}, {C}_{k}\right).
\end{equation}
In the decoder, let $B_k$ be the output of the first sublayer. Similar to the encoder, the second sublayer is the context attention:
\vspace{-1mm}
\begin{equation}
	{N}_{k}=\operatorname{MultiHead}\left({B}_{k}, {C}_{k}, {C}_{k}\right).
	\vspace{-1mm}
\end{equation}
We use a gating method to regulate the contextual information:
\vspace{-2mm}
\begin{equation}
	\operatorname{Gating}(C_k)=\lambda {C_k}+(I-\lambda) N_k,
\end{equation}
where $I$ being a unit vector, $\lambda$ is the gating vector given by:
\vspace{-1mm}
\begin{equation}
	\lambda=sigmoid\left({W}_{i} {C_k}+{W}_{s} N_k \right),
	\vspace{-1mm}
\end{equation}
where ${W}_{i}$ and ${W}_{s}$ are trainable parameters.
$\lambda$ represents the learned gates applied to dimensions of $C_k$ to weight the importance.
The target phoneme $y_k$ is predicted using the softmax:
\begin{equation}
	P\left(y_k \mid {x}_{k}, {C}_{k}; \theta\right) \propto \exp \left({W}_o \times {T}_k\right),
\end{equation}
where ${W}_o \in \mathbb{R}^{\left|\mathcal{V}_{y}\right| \times d}$ is a model parameter, $\mathcal{V}_{y}$ is the phoneme vocabulary.
$T_k \in \mathbb{R}^{d \times 1}$ is a column vector for predicting $y_k$.
The context may provide auxiliary information (e.g., part-of-speech, word category, word frequency) for the noisy words, thus is helping to "correct" the noise in the latent space.

\subsection{Robust training}
Joint training the contextual information may increase the influence in an uncontrolled way. To avoid this, we propose a two-step training strategy that uses a clean word-level corpus $D_w$ along with the noisy sentence-level corpus $D_s$. We divide model parameters into two subsets: ${\theta}_{w}$ (highlighted in blue in Fig. 3) and ${\theta}_{s}$ is newly-introduced (in yellow).
In the first step, ${\theta}_{w}$ are optimized on the combined corpus $ D_{w} \cup D_{s} $:
\vspace{-1mm}
\begin{equation}
	\tilde{\theta}_{w}=\underset{\theta_{w}}{\operatorname{argmax}} \sum_{\langle{x}, {y}\rangle \in D_{w} \cup D_{s}} \log P\left({y} \mid {x} ; \theta_{w}\right).
	\vspace{-1mm}
\end{equation}
In the second step, ${\theta}_{s}$ are optimized on the $D_s$ only:
\vspace{-1mm}
\begin{equation}
	\tilde{\theta}_{s}=\underset{\theta_{s}}{\operatorname{argmax}} \sum_{\langle{X}, {Y}\rangle \in D_{s}} \log P\left({y} \mid {x} ; \tilde{\theta}_{w}, \theta_{s}\right).
	\vspace{-1mm}
\end{equation}
This is similar to the pre-training. The major difference is that ours fixes $\tilde{\theta}_{w}$ when optimizing ${\theta}_{s}$ to prevent overfitting to the noisy $D_s$ and stablize the training by the way.

\section{Experiments and Results}
\label{exp}
\subsection{Data Preparations and Model Details}
The CMUdict 0.7b~\cite{cmu} are adopted as $D_w$.
$D_s$ comes from CCOHA~\cite{cc}, which is clean and of large diversity in words.
The test sets are two aspects: (1) \textbf{Two} dictionary-based benchmarks: CMUdict and NetTalk~\cite{net}; (2) \textbf{Three} real-world noisy sets, which are same in Sec.~\ref{sec:real}.
We extract the noisy-corrected text pairs, and convert the corrected texts into ground-truth phonemes.
The transformer has 4 encoder-decoder layers and 4 attention heads.
The embedding size is 128 for graphemes and phonemes, 512 for word embeddings in the context.
During inference, the beam search is used with size of 4.

\vspace{-1mm}
\subsection{Main results}
Through all hyperparameters tried, we report the best results in Table~\ref{ro}.
On the dict-based clean test sets, our models all outperform the baseline.
\textbf{adv} achieves the most improvement of -2.25 \% WER on CMUdict and -3.21 \% WER on NetTalk.
This proves our method can improve the robustness in the clean scenario.
Previous methods perform well on the clean data but suffer from a great performance drop on noisy testsets.
r-G2P significantly outperforms in noisy scenarios.
\textbf{syn} obtains the best performance of -9.09 \% WER compared with the baseline.
We also compare with first correcting noisy texts by Microsoft Bing Spell Check~\cite{bing}, then converting by the baseline.
Ours conversion accuracy is also higher.
Considering the cost to integrate a spell corrector before G2P models, ours is a unified model which is more efficient and accurate.

\begin{table*}[t]
	\centering
	\caption{Comparisons on various test sets. Numbers form in PER (\%) / WER (\%). Note that~\cite{bil} has no official code.}
	\label{ro}			
	\resizebox{1.0\linewidth}{18mm}{
		\begin{tabular}{c|cc|ccc}
			\toprule[2pt]
			\multirow{2}{*}{\textbf{Method}} & \multicolumn{2}{c|}{\textbf{Dict-based Benchmarks}} &
			\multicolumn{3}{c}{\textbf{Real-world Sources}} \\ 
			& \textbf{CMUdict} & \textbf{NetTalk} & \textbf{News Crawl} & \textbf{TED Talks} & \textbf{ICDAR} \\ \midrule[1pt]
			\textbf{Encoder CNN, decoder Bi-LSTM (fifth model)~\cite{bil}} & \textbf{4.81} / 25.13 &  5.69 / 30.10  & --- & --- & --- \\ 
			~~~~~\textbf{CNN with NSGD~\cite{nsgd}}~~~~~ & ~~~5.58 / 24.10~~~ &  ~~~6.78 / 28.45~~~ & ~~~~~~12.43 / 43.82~~~~~~ & ~~~~~~15.12 / 45.96~~~~~~ & ~~~~~~19.36 / 49.05~~~~~~ \\
			\textbf{Encoder-decoder + global attn~\cite{bi}} & 5.04 / 21.69 & 7.14 / 29.20 & 16.17 / 44.57 & 16.34 / 47.20 & 18.28 / 45.43 \\
			\midrule[1pt]
			\textbf{Transformer 4x4}  & 5.23 / 22.10 & 6.87 / 29.82 & 11.10 / 42.83 & 15.90 / 44.56 & 16.01 / 42.06 \\
			\textbf{Bing + Transformer 4x4} & --- & --- & 8.78 / \textbf{32.61} & 10.29 / 36.94 & 12.15 / 37.52   \\ \midrule[1pt]
			\textbf{r-G2P (nat)} & 5.22 / 20.14 & 6.64 / 28.85 & 9.94 / 33.45 & \textbf{8.16} / 36.25 & 11.58 / 36.94 \\
			\textbf{r-G2P (syn)}  & 5.09 / 21.67 & 6.68 / 29.13 & \textbf{8.61} / 32.76  & 8.64 / \textbf{35.06} & \textbf{10.39 / 34.35} \\
			\textbf{r-G2P (adv)} & 4.84 / \textbf{19.85} & \textbf{5.34} / \textbf{26.61} & 10.31 / 36.53 & 9.65 / 40.72 & 13.46 / 39.42   \\ \bottomrule[2pt]
	\end{tabular}}
	\vspace{-2mm}
\end{table*}

\vspace{-1mm}
\subsection{Effect of noise ratio}
\begin{figure}[h]
	\centering
	\subfigure[\textbf{nat}]{
		\begin{minipage}{0.5\linewidth}
			\centering
			\includegraphics[width=1.6in]{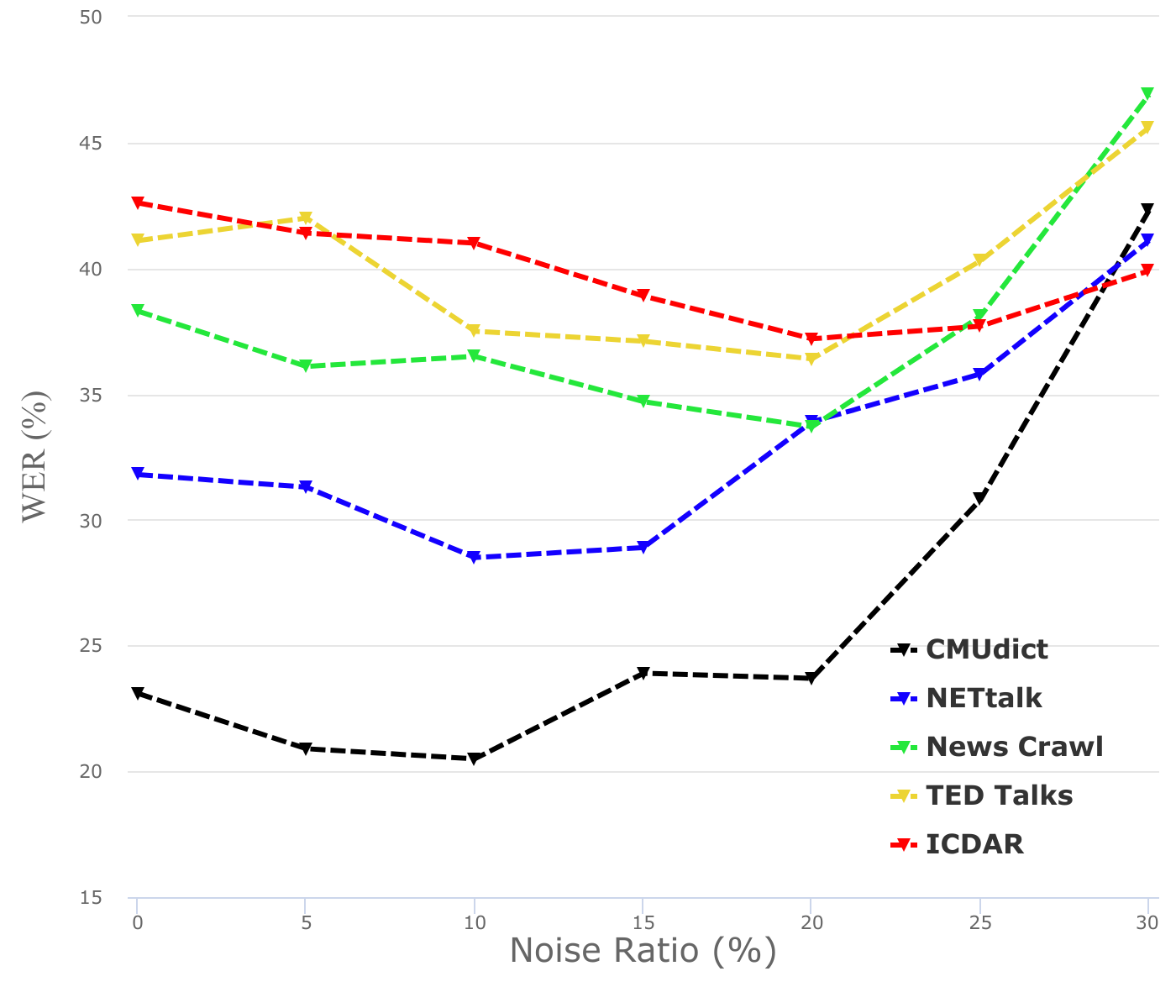}
		\end{minipage}%
	}%
	\subfigure[\textbf{syn}]{
		\begin{minipage}{0.5\linewidth}
			\centering
			\includegraphics[width=1.6in]{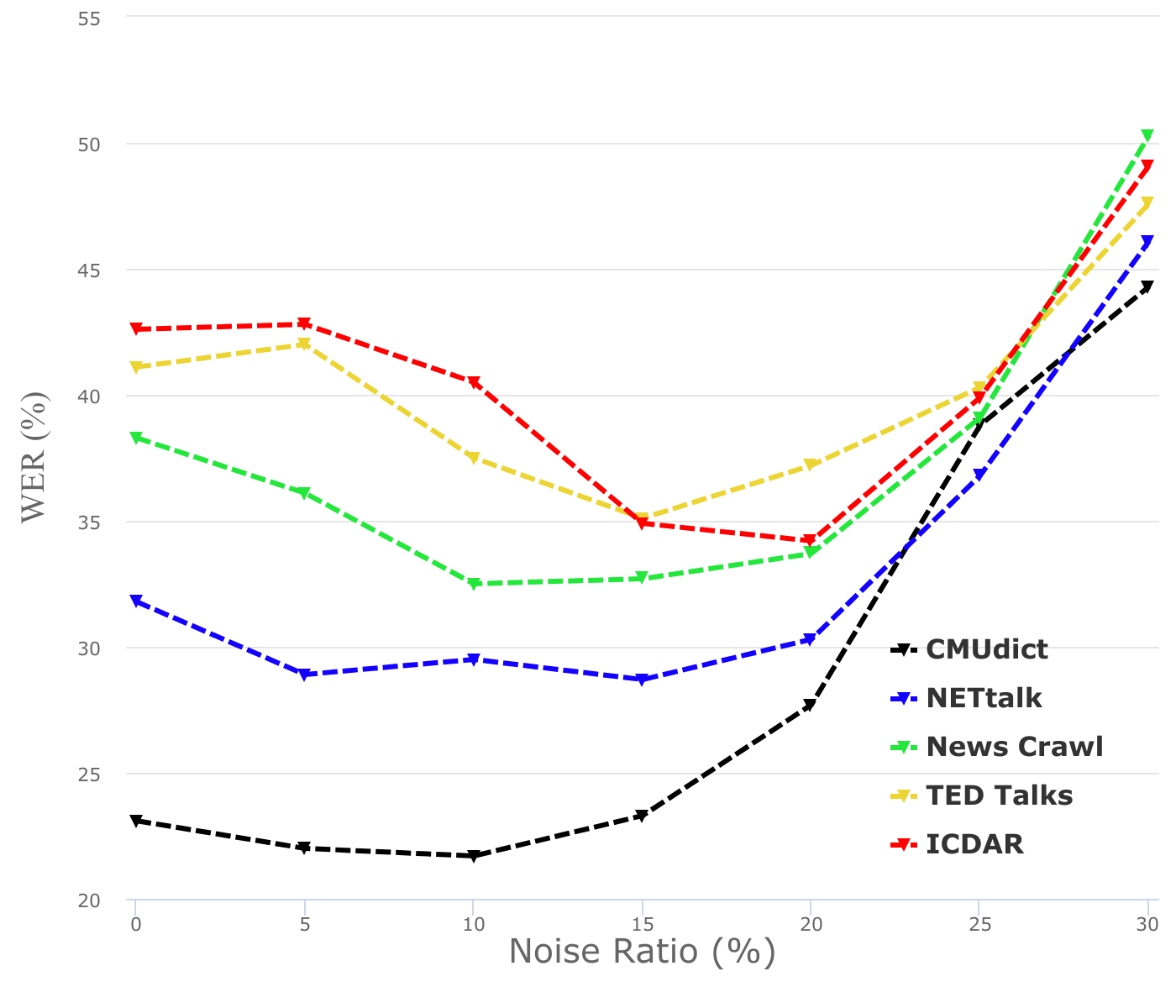}
		\end{minipage}%
	}%
	\vspace{-2mm}
	\caption{ Varying noise ratio $p$ results in model performances. }
	\label{fig:noise}
\end{figure}

We setup noise ratio $p$ in the whole sentence-level corpus.
Fig.~\ref{fig:noise} shows the effect of $p$.
It is clear that introducing noise indeed enhances performance on not only the real-world noisy data but also the dict-based clean data.
A moderate amount of noise would enhance the robustness, while the moderate amounts are not fixed for each dataset depending on the noise distributions.
The noisy sets obviously adjust to a higher noise ratio, which is too high for clean sets.
This also indicates that the different noise introducing methods have different impacts.
With the same $p$, \textbf{syn} makes a stronger impact than \textbf{nat}.

\vspace{-1mm}

\subsection{Effect of context length}
We investigate the effect of context length and report the average WER here. $p = 0.2$ are set as a stationary point. As shown in Table~\ref{len}, incorporating the contextual information greatly enhances the anti-noise capability.
Setting $l = 1$ or $l = 2$ as sentence-level context achieves the best performance. Using more words does not bring further improvement and increases computational cost.
It's worth noticing that incorporating the contextual information has almost no contribution to \textbf{adv}.
We assume that this kind of low-level feature is not helpful for perturbations in the high-dimensional hidden space.  
Results prove that the contextual information and robust training can cooperate with each other to improve the performance further.

\begin{table}[h]
	\centering
	\vspace{-2mm}
	\caption{ Context length $l$ results in WER (\%). \textbf{Ro.} denotes the robust training. $l = 0$ means no context integrated. }
	\label{len}
	\resizebox{1.0\linewidth}{14mm}{
		\begin{tabular}{c|c|c|c|c|c}
			\toprule
			 & \textbf{Ro.} & \textbf{$l = 0$} & \textbf{$l = 1$} & \textbf{$l = 2$} & \textbf{$l = 3$} \\ \midrule
			\multirow{2}{*}{\textbf{r-G2P (nat)}} & --  & 41.29   & 38.74 & 37.08  & 36.61   \\
			 & \checkmark  & 40.37   &  \textbf{35.37}   &  35.52  &  35.46    \\ \midrule
			\multirow{2}{*}{~~~~\textbf{r-G2P (syn)}~~~~} & -- & ~~~40.46~~~ & ~~~39.54~~~ & ~~~36.42~~~ & ~~~36.03~~~    \\
			& \checkmark & 39.64     &  36.78  & \textbf{35.09}     &  35.28  \\ \midrule
			\multirow{2}{*}{\textbf{r-G2P (adv)}} & --  & 40.15    & 41.23     & 40.98      & 41.39      \\
			& \checkmark  & 38.23 & 38.51 &    \textbf{38.14}            &    38.38            \\ \bottomrule
		\end{tabular}}
	\vspace{-3mm}
\end{table}

\subsection{Case study}
Addressing out-of-vocabulary (OOV) words (i.e., abbreviations, foreign names) is a major goal for G2P models.
We extract some realistic examples from Wiktionary in Table 3. 
Compared with the baseline, ours can generate more accurate predictions.
On the other hand, CMUdict is crafted manually and updated consistently, hence prone to annotation errors.
The misspelled word "commerical" is annotated with the same label as "commercial". Even for this non-existent word, ours outputs quite a convincing pronunciation.

\begin{table}[h]
	\centering
	\label{case}
	\vspace{-3mm}
	\caption{Examples converted by the baseline Transformer and ours. \textbf{GT} denotes the ground truth, where red being wrong.}
	\resizebox{1.0\linewidth}{11mm}{
		\begin{tabular}{cccccc}
			 & \textbf{Name}  & \textbf{Abbr.} & \textbf{Compound}  & \multicolumn{2}{c}{\textbf{Wrong GT}} \\ \toprule
			\multicolumn{1}{c|}{\textbf{Word}} & Xochitl & ASAP & coathanger & commerical & honest \\ \midrule
			\multicolumn{1}{c|}{\textbf{GT }} & \textesh o\textlengthmark t\textesh itl  & e\i s\ae p & ko\textupsilon t h\ae \ng \textrevepsilon  & k\textschwa m\textrevepsilon \textlengthmark \textcolor{red}{\textesh}\textschwa l & \textscripta \textcolor{red}{\textlengthmark}n\textcolor{red}{\textschwa}st \\
			\multicolumn{1}{c|}{\textbf{Trans}} & z\textscripta \textlengthmark k\textschwa t\textschwa l & e\i \textepsilon se\i pi\textlengthmark & ko\textupsilon \texttheta \ae g\textrevepsilon & k\textschwa m\textrevepsilon \textlengthmark \textesh \textschwa l & \textscripta \textlengthmark n\textschwa st \\
			\multicolumn{1}{l|}{\textbf{r-G2P}} & ko\textlengthmark t\textesh itl & e\i s\ae p & ko\textupsilon t h\ae \ng \textrevepsilon & k\textschwa m\textrevepsilon \textlengthmark r\i k\textschwa l & \textscripta n\i st \\ \bottomrule
		\end{tabular}}
	\vspace{-1mm}
\end{table}

\begin{figure}[h]
	\centering
	\includegraphics[width=0.5\linewidth]{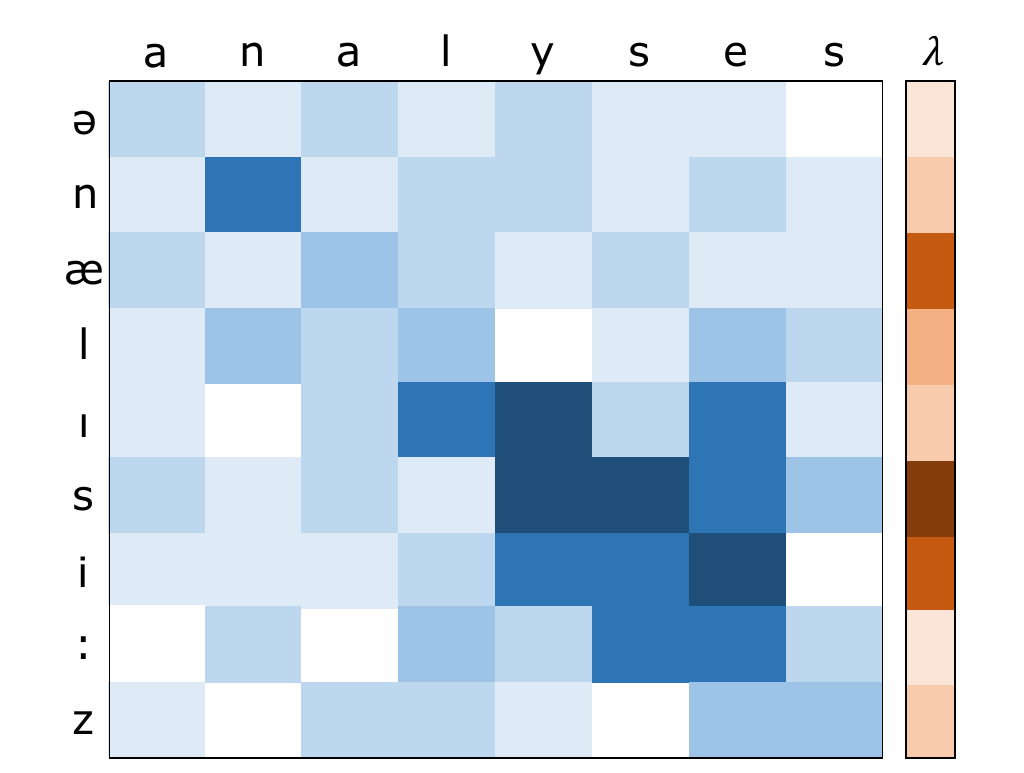}
	\caption{ Visualizations of G-P alignments and the gating $\lambda$.
		Darker color means greater alignment weights and gated scales.
		The context is " ... some chemical analyses of the ... ".}
	\label{fig:map}
\end{figure}

Another issue is the homograph disambiguation, where a word pronounced differently depending on the context.
For example, "analyses" is both the third-person singular form of "analyse" ( [\ae n\textschwa l\textscripta \i z\i z] ) and the plural of "analysis" ( [\textschwa n\ae l\i si\textlengthmark z] ).
We extract the encoder-decoder attention map and visualize it in Fig.~\ref{fig:map}. Ours took advantage of the contextual information, which mainly focused on the hard-to-distinguish phonemes, and converted this word correctly.

\vspace{-3mm}
\section{Conclusion}
In this paper, we mainly raise the issue of G2P model's robustness on the noisy words since it has never been explored.
We first confirm its vulnerability and statistically analyze the conversion failures caused by the noise. Then, we propose three controlled noise introducing methods to the training data. By incorporating the contextual information and the robust training process, we substantially mitigate the noise effect and achieve a robust G2P model. Experimental results show that the r-G2P significantly outperforms previous methods both on the dict-based benchmarks and in real-world scenarios.


\section{ACKNOWLEDGMENT}
This paper is supported by the Key Research and Development Program of Guangdong Province under grant No. 2021B0101400003 and the National Key Research and Development Program of China under grant No. 2018YFB0204403. Corresponding author is Jianzong Wang from Ping An Technology (Shenzhen) Co., Ltd (jzwang@188.com).


\bibliographystyle{IEEEbib}
\bibliography{strings,refs}

\begin{thebibliography}{10}

\bibitem{1}
Zach Ryan and Mans Hulden,
\newblock ``Data augmentation for transformer-based {G2P},''
\newblock in {\em Proceedings of the 17th {SIGMORPHON} Workshop on
  Computational Research in Phonetics, Phonology, and Morphology, {SIGMORPHON}
  2020, Online, July 10, 2020}. pp. 184--188, Association for Computational
  Linguistics.

\bibitem{2}
Bradley Hauer, Amir~Ahmad Habibi, Arnob Mallik, and Grzegorz Kondrak,
\newblock ``Low-resource {G2P} and {P2G} conversion with synthetic training
  data,''
\newblock in {\em Proceedings of the 17th {SIGMORPHON} Workshop on
  Computational Research in Phonetics, Phonology, and Morphology, {SIGMORPHON}
  2020, Online, July 10, 2020}. pp. 117--122, Association for Computational
  Linguistics.

\bibitem{3}
Ben Peters and Andr{\'{e}} F.~T. Martins,
\newblock ``One-size-fits-all multilingual models,''
\newblock in {\em Proceedings of the 17th {SIGMORPHON} Workshop on
  Computational Research in Phonetics, Phonology, and Morphology, {SIGMORPHON}
  2020, Online, July 10, 2020}. pp. 63--69, Association for Computational
  Linguistics.

\bibitem{4}
Zhenhou Hong, Jianzong Wang, Xiaoyang Qu, Jie Liu, Chendong Zhao, and Jing
  Xiao,
\newblock ``{Federated Learning with Dynamic Transformer for Text to Speech},''
\newblock in {\em Proc. Interspeech 2021}, 2021, pp. 3590--3594.

\bibitem{trans}
Sevinj Yolchuyeva, G{\'e}za N{\'e}meth, and B{\'a}lint Gyires-T{\'o}th,
\newblock ``Transformer based grapheme-to-phoneme conversion,''
\newblock {\em Proc. Interspeech 2019}, pp. 2095--2099.

\bibitem{ic1}
Yonghe Wang, Feilong Bao, Hui Zhang, and Guanglai Gao,
\newblock ``Joint alignment learning-attention based model for
  grapheme-to-phoneme conversion,''
\newblock in {\em IEEE International Conference on Acoustics, Speech and Signal
  Processing (ICASSP)}. IEEE, 2021, pp. 7788--7792.

\bibitem{ic2}
MAC Dang-Khoa, NGUYEN, and Kim-Anh NGUYEN,
\newblock ``How to make text-to-speech system pronounce" voldemort": an
  experimental approach of foreign word phonemization in vietnamese,''
\newblock in {\em IEEE International Conference on Acoustics, Speech and Signal
  Processing (ICASSP)}. IEEE, 2021, pp. 6483--6487.

\bibitem{adn}
Haipeng Sun, Rui Wang, and Tiejun Zhao,
\newblock ``Robust unsupervised neural machine translation with adversarial
  denoising training,''
\newblock in {\em Proceedings of the 28th International Conference on
  Computational Linguistics, {COLING} 2020}. 2020, pp. 4239--4250,
  International Committee on Computational Linguistics.

\bibitem{news}
Lo{\"\i}c Barrault and Ond{\v{r}}ej Bojar,
\newblock ``Findings of the 2019 conference on machine translation ({WMT}19),''
\newblock in {\em Proceedings of the Fourth Conference on Machine Translation
  (Volume 2: Shared Task Papers, Day 1)}, Florence, Italy, Aug. 2019, pp.
  1--61, Association for Computational Linguistics.

\bibitem{ted}
Mattia~A Di~Gangi, Cattoni, and Marco Turchi,
\newblock ``Must-c: a multilingual speech translation corpus,''
\newblock in {\em 2019 Conference of the North American Chapter of the
  Association for Computational Linguistics}. Association for Computational
  Linguistics, 2019, pp. 2012--2017.

\bibitem{OCR}
Christophe Rigaud, Antoine Doucet, Mickaël Coustaty, and Jean-Philippe Moreux,
\newblock ``Icdar 2019 competition on post-ocr text correction,''
\newblock in {\em 2019 International Conference on Document Analysis and
  Recognition (ICDAR)}, 2019, pp. 1588--1593.

\bibitem{pedia}
``Wikipedia,''
  \url{https://en.wikipedia.org/wiki/Wikipedia:Lists_of_common_misspellings}.

\bibitem{cvv}
Dominic~W Massaro and Michael~M Cohen,
\newblock ``Phonological context in speech perception,''
\newblock {\em Perception \& psychophysics}, vol. 34, no. 4, pp. 338--348,
  1983.

\bibitem{adv}
Takeru Miyato, Andrew~M. Dai, and Ian~J. Goodfellow,
\newblock ``Adversarial training methods for semi-supervised text
  classification,''
\newblock in {\em 5th International Conference on Learning Representations,
  {ICLR} 2017, Toulon, France, April 24-26, 2017, Conference Track
  Proceedings}, 2017.

\bibitem{cmu}
``Cmu pronouncing dictionary,''
  \url{http://www.speech.cs.cmu.edu/cgi-bin/cmudict}.

\bibitem{cc}
Reem Alatrash, Dominik Schlechtweg, Jonas Kuhn, and Sabine~Schulte im~Walde,
\newblock ``Ccoha: Clean corpus of historical american english,''
\newblock in {\em Proceedings of The 12th Language Resources and Evaluation
  Conference}, 2020, pp. 6958--6966.

\bibitem{net}
``\text{NetTalk},''
  \url{https://archive.ics.uci.edu/ml/datasets/Connectionist+Bench+(Nettalk+Corpus)}.

\bibitem{bing}
``\text{Bing Spell Check},''
  \url{https://azure.microsoft.com/en-in/services/cognitive-services/spell-check}.

\bibitem{bil}
Sevinj Yolchuyeva, G{\'e}za N{\'e}meth, and B{\'a}lint Gyires-T{\'o}th,
\newblock ``Grapheme-to-phoneme conversion with convolutional neural
  networks,''
\newblock {\em Applied Sciences}, vol. 9, no. 6, pp. 1143, 2019.

\bibitem{nsgd}
Moon-Jung Chae and Kyubyong Park,
\newblock ``Convolutional sequence to sequence model with non-sequential greedy
  decoding for grapheme to phoneme conversion,''
\newblock in {\em IEEE International Conference on Acoustics, Speech and Signal
  Processing (ICASSP)}. IEEE, 2018, pp. 2486--2490.

\bibitem{bi}
Shubham Toshniwal and Karen Livescu,
\newblock ``Jointly learning to align and convert graphemes to phonemes with
  neural attention models,''
\newblock in {\em 2016 IEEE Spoken Language Technology Workshop (SLT)}, 2016,
  pp. 76--82.

\end{thebibliography}

\end{document}